\documentclass[12pt,preprint]{aastex}

\shorttitle{Bulge formation in responding cuspy dark matter halo}
\shortauthors{Fu, Liu, Huang \& Deng }

\begin{document}


\title{Bulge formation from SSCs in a responding cuspy dark matter halo}


\author{Yan-Ning Fu}
\affil{Purple Mountain Observatory, Chinese Academy of Sciences,
Nanjing 210008, China} \affil{National Observatories, Chinese
Academy of Sciences, Beijing 100012, China} \email {fyn@pmo.ac.cn}

\author{Wen-Hao Liu} \affil{Department of Astronomy and Physics, University of California, Irvine,
California 92617, USA} \email {wenhaol@uci.edu}

\author{Jie-Hao Huang}
\affil{Department of Astronomy, Nanjing University, Nanjing 210093,
China} \email {jhh@nju.edu.cn}

\and

\author{Zu-Gan Deng} \affil{College of Physical Science, Graduate School of the Chinese Academy of Sciences, Beijing 100049,
China} \email {dzg@vega.pku.bac.ac.cn}




\begin{abstract}
We simulate the bulge formation in very late-type dwarf galaxies
from circumnuclear super star clusters (SSCs) moving in a responding
cuspy dark matter halo (DMH). The simulations show that (1) the
response of DMH to sinking of SSCs is detectable only in the region
interior to about 200 pc. The mean logarithmic slope of the
responding DM density profile over that area displays two different
phases: the very early descent followed by ascent till approaching
to 1.2 at the age of 2 Gyrs. (2) the detectable feedbacks of the DMH
response on the bulge formation turned out to be very small, in the
sense that the formed bulges and their paired nuclear cusps in the
fixed and the responding DMH are basically the same, both are
consistent with $HST$ observations. (3) the yielded mass correlation
of bulges to their nuclear (stellar) cusps and the time evolution of
cusps' mass are accordance with recent findings on relevant
relations. In combination with the consistent effective radii of
nuclear cusps with observed quantities of nuclear clusters, we
believe that the bulge formation scenario that we proposed could be
a very promising mechanism to form nuclear clusters.
\end{abstract}




\section{Introduction}

In our previous works (Fu, Huang, \& Deng 2003 (FHD03), Huang, Deng, \&
Fu 2003 (HDF03)), we managed to construct a set of models for
simulating the dynamical evolution of super star clusters (SSCs) in
dark matter halo (DMH). The simulations with vastly different
settings consistently yielded bulges similar to the observational
ones. Two main similarities are (1) the derived surface density can
be well fitted by an exponential profile with an additional nuclear
cusp, which is consistent with {\it Hubble Space Telescope (HST)}
observations (Carollo 1999); (2) There is a causal connection
between the masses of the paired bulge and nuclear cusp, which is
compatible with the observational evidence about a tight luminosity
correlation between the paired two (Balcells et al. 2003).

However, the DMH response to SSCs sinkage was not considered in both
FHD03 and HDF03. The DMH response, in addition to its feedback on
the bulge formation, is also interesting in the context of the inner
slope of the DMH profile. As is well known, numerical simulations
show that the density profile of virialized $\mathrm{\Lambda}$CDM
DMHs have central cusps (e.g. Navarro et al. 1997; Moore et al.,
1999; Jing \& Suto 2002; Navarro et al. 2004). However, high
resolution observations of dwarfs and low surface brightness
galaxies (de Blok et al., 2003; Swaters et al., 2003; Simon et al.
2005) indicate that the Burkert profile (Burkert 1995) with constant
density core instead of the cuspy NFW one is more suitable to these
galaxies. This apparent contradiction has drawn much concerns in
astronomical communities. While some groups were checking the
conflict by performing refined simulations or conducting higher
resolution observations  (e.g. Power et al. 2003; Navarro et al.,
2004; Simon et al.,  2005 and the references therein), the others
were trying to find a solution to this problem for CDM by
considering additional physical process involving luminous matter.
In particular, El-Zant et al. (2001, 2004) showed that the DM
distribution flattens when gas clumps in a galaxy or galaxies in
clusters spiral inward.

On the other hand, Kazantzidis et al. (2004) have shown that the DMH
initialized using the local Maxwellian approximation can result in
spurious evolution. It was indeed what we faced with during our
early investigation on bulge formation in a responding DMH to
sinking SSCs. We found false evolution of DMH, due to the fact that
the DMH is not in equilibrium under the local Maxwellian
approximation. To ensure real response of host DMH to sinking SSCs,
we'll start with building an equilibrium DMH using exact
distribution function (Kazantzdis et al., 2004, Liu et al. 2005) in
this work, followed by considering the DMH response to SSCs sinkage,
namely, the effects of gravitational contraction and heating
expansion caused by spiraling inward and the transferring orbital
energy of SSCs. Taking the above two effects into consideration, we
simulate the dynamical evolution of SSCs in an initially cuspy DMH,
as a more sophisticated model of bulge formation in very late-type
galaxies. The properties of the formed bulges and their paired
nuclear (stellar) cusps will be compared with, in addition to the
ones already mentioned, the most recent observations (Rossa et al.
2006), where the authors found a correlation between nuclear cluster
mass and the bulge luminosity for their 40 sample spirals.

The structure of this paper is as follows. In section 2, we describe
our models. Sections 3 and 4 are devoted respectively to presenting
and discussing our simulation results about DMH evolution and bulge
formation. Conclusions are given in section 5.

\section{Models}

In this work, the bulge formations in a set of 40 sample galaxies
are simulated. Apart from the DMH-related models, which will be
described in the following subsections, the other settings are the
same as those of HDF03 and listed in Table 1.  The basic of these
settings are summarized below for completeness.

The radial distribution of SSCs is obtained from the projected
radial number distribution of  bright star-forming complexes in a
sample of  very late-type galaxies (Parodi \& Binggeli  2003),
$R{\hspace*{0.6mm}} \exp(-R/R_{\rm l})$ with a median scale length
of $R_{\rm l} \approx 0.6${\hspace*{0.6mm}}kpc and a median number
of bright star-forming complexes of 17, respectively. These SSCs are
all modeled as a truncated isothermal sphere. The initial mass
function of SSCs is assumed to be a log-normal mass function,
written as

\begin{equation}
\label{SSCmassspectrum} Log_{10}(\frac{M}{M_{\odot}}) \sim  N(
\exp(mean)=2 \times 106 M_{\odot}, variance=0.08)
\end{equation}

We assume that the stellar mass outside a sphere of radius $R_t$,
which corresponds to the instantaneous Hill stability region around
the SSC center, will be stripped. On average, the stars stripped
when the SSC goes from $r$ to $r-dr$ contribute to a region radially
bounded by $r+R_t(r)$ and $r-dr-R_t(r-dr)$. In our simulations,
tidal  stripping is not allowed when the mass of the stripped SSC is
less than $1 M_{\odot}$.

\subsection{Numerical DMH}

\subsubsection{Simulation methods}
In order to consider the effects of the interaction between the DMH
and the SSCs on the bulge formation, we will use a particle
representation of the DMH. When one uses limited number of particles
to represent a gravitational system and perform a usual N-body
simulation, he in fact assumes that each representing particle is
actually a representative of the physical ones with similar
3-dimensional velocity in its 3-dimensional neighborhood. Since our
model is assumed to be spherically symmetric, each particle in our
simulation is reasonably used to represent the physical particles
with similar radial and transverse velocities in a shell. It can be
understood that the radial position of a shell corresponds to the
3-dimensional position of the above-mentioned neighborhood in the
usual N-body simulation. In doing the numerical integration, one can
follow El-Zant, Shlosman, \& Hoffman (2001) and assume that the
gravitation on a given particle at $\mathbf{r}$ is
$-(GM(r)/r^3)\mathbf{r}$, where $\mathbf{r}$ is the position vector
starting from the center, $r=|\mathbf{r}|$, and $M(r)$ the total
mass interior to $r$. This force model ensures automatic maintenance
of the spherical symmetry. Based on this perspective (e.g., Henon,
1971; El-Zant , Shlosman, \& Hoffman  2001), we believe that the way
we use to integrate the numerical DMH is well adapted to nearly
spherically symmetric systems like ours.

One basic requirement for this representation is that it must be
statistically meaningful at $5\ pc$ from the center, so that the
simulations could be compared with the aforementioned observations
with linear resolution of about $5\ pc$. Simple calculation shows
that, even if this requirement is as weak as to have about 10
particles in the central globe of radius $5\ pc$, one needs about
$10^{8}$ particles for representing the whole DMH interior to
$r_{200}$. This would make it extremely time-consuming to complete
our simulations.

To work around this difficulty, we neglect the DMH response outside
$1\ kpc$. Indeed, our preliminary simulations show that the DMH
response is hardly detectable at locations more than about $500\ pc$
from the center (Fu, Huang, \& Deng 2004 (FHD04)). Physically, this
phenomenon has a connection with a reasonable general belief, that
is, a low-mass intruder shouldn't induce global evolution of a
high-mass intruded system. In our case, the DMH mass of $3.4 \times
10^8 M_{\odot}$ initially confined to the central globe of radius
$1\ kpc$ is more than 10 times larger than any single SSC mass. In
fact, the SSC mass initially inside 1kpc in our sample galaxies
never exceeds $3.7 \times 10^7 M_{\odot}$, about ten times smaller
than the dark matter mass in the same region. Also, as can be
inferred from our previous simulations and verified ex post facto by
the present one, only a small number of SSCs will spiral down to
locations less than $1\ kpc$ from the galactic center on the time
scale of bulge formation. Therefore, the simplifying assumption that
there's no DMH response outside $1\ kpc$ shouldn't have any severe
influence on the simulation results.

A practical problem of using full N-body simulation for our
investigation is that one is not able to use particles with small
enough mass to represent the dark matter halo. This means that, when
masses of SSCs become not much massive than the representing
particle of dark matter halo due to tidal stripping, the effect of
dynamical friction cannot be correctly reproduced for free in
practice. For example, if we set $ 2 \times 10^7$ particles, about
10 solar mass of each representing particle,  inside of 1 kpc for
our simulations, the effect of dynamical friction would not be
properly reproduced for stripped SSCs  with less than about 100
solar mass. For a large number of SSCs, however,  the mass of
remaining SSCs at around 5 pc from the center would  be a few tens
of solar mass or less. Therefore, $ 10^8 $ or more particles are
needed in order to properly include the DMH heating in the inner 5pc
globe for free, as is required for the proposed bulge formation
processes to be simulated with the required resolution of 5 pc.

It follows that even using a full N-body code for our investigation,
one has to add the dynamical friction by hand. On the other hand, a
number of investigations performed with N-body simulations have
provided the basis for taking the semi-analytical approach (e.g.
Cora, Muzzio, \& Vergne 1997; Velazquez \& White, 1999). These
studies indicated that ``the numerical (i.e. full N-body) results of
low-mass satellites showed very good agreement with theoretical
prediction obtained from straightforward application of
Chandraehkhar dynamical friction equation" (Cora, Muzzio, \& Vergne
1997), and they found ``Chandraehkhar's dynamical friction formula
works well provided a suitable value is chosen for the Coulomb
logarithm and the satellite mass is taken to be the mass still bound
to the satellite at each moment" (Velazquez \& White, 1999). In view
of this, the Coulomb logarithm in our simulations is calculated each
time the formula is applied, according to the relevant updated
quantities, and, by taking account the stripping effect, only the
mass still bound to the SSC is used.

But we must point out that the above-mentioned numerical experiments
didn't follow the satellite all the way in towards the center of its
host galaxy, where the semi-analytical equation (2) could become
invalid because the size of the satellite is not always adequately
small as compared to the scale length of the background density
variation. While this possible inadequacy could have some influences
on the bulge formation process, the overall influence should not be
severe because, on its way towards the center, an SSC also becomes
smaller and smaller due to tidal stripping. In fact, most SSCs will
be destroyed before they are close to the center, and the remaining
ones are very compact.

Apart from computation time, the two methods, N-body and
semi-analytical approaches, are expected to be equivalent in regard
to dealing with the dynamical friction, except for the unfavorable
region very close to the galactic center. Considering the
time-saving advantage on taking the semi-analytical experiment as
described in this paper, especially for simulating at least several
tens of sample galaxies, we tend to adopt the later approach for our
study.

However, it is surely important to reveal some potential
inadequacies, induced by the semi-analytical treatment of dynamical
friction, as well as of other physical processes to be introduced in
the following subsections, by simulating the proposed bulge
formation processes with a full N-body code, such as GADGET, when
sufficiently powerful computer facility is available. This will help
one to decide to what extent the semi-analytical approximations can
be accepted with full confidence.


\subsubsection{Reflecting sphere}

Under the above assumption, we model our DMH as follows. The DMH
content interior to the spherical surface of radius $1.1\ kpc$ is
represented by a system of $10^6$ particles. These particles are
randomly generated according to the isotropic stable distribution
function corresponding to the NFW density profile but with an outer
exponential cutoff (Kazantzidis et al. 2004; Liu et al. 2005).
During the numerical integration of the particle system, a
reflecting boundary condition will be applied on the above-mentioned
spherical surface. To be specific, whenever an outgoing particle
reaches this surface, its velocity will be reversed.

In other words, the inflow halo mass across the sphere of radius
1.1kpc is accounted for by the reflected particles at the inner
boundary of the same sphere. This is based on the assumption that
this sphere is situated in the nearly stable region of the halo,
which, in particular, means that each inflow halo particle should be
compensated by an outflow one with the same velocity. In practice,
the boundary condition cannot be applied exactly on the sphere,
there are inward particles lie slightly outside 1.1kpc. These
particles are allowed to go inside freely.

For a given sample particle system, the part inside $1\
 kpc$, will be referred to as a numerical
DMH, of which the particle positions will be used in calculating the
gravitation of the DMH at distances interior to $1\ kpc$. For
particles outside $1\ kpc$, the gravitation will be calculated by
using the analytical NFW model. This is because we discuss the
evolution of the DMH profile only inside 1kpc, outside which the
halo is assumed to be in equilibrium.

\begin{figure*} 
   \vspace{3cm}
   \hspace{0 cm}\includegraphics[width=18cm]{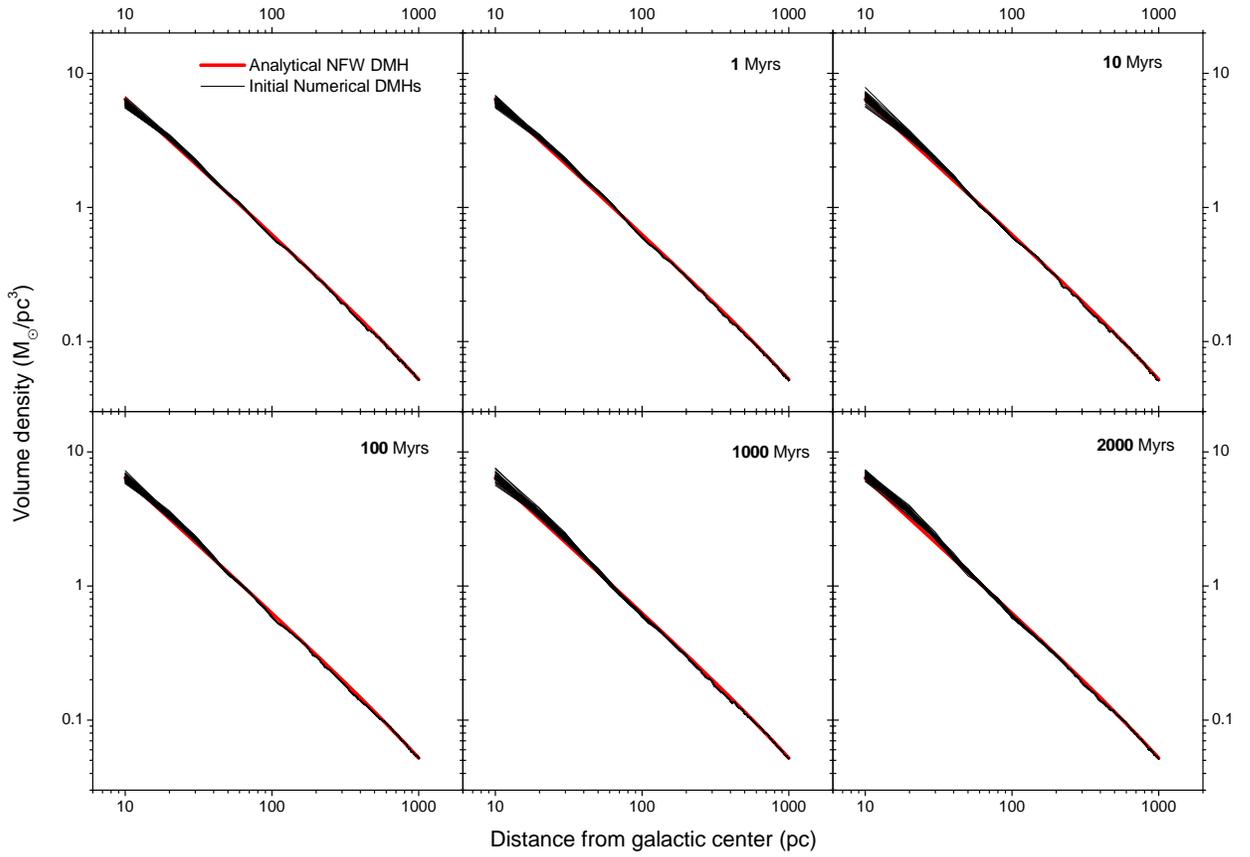}
      \vspace{-6cm}\caption{Volume density profiles of the numerical DMHs in the
      absence of SSCs moving in them (we will name it ``unpolluted DMHs'' in the
      next figures).
      } \label{nhd}
   \end{figure*}

To see to what extent our numerical DMHs can be claimed as stable
systems, we integrate them in the absence of SSCs.  The volume
density profiles of the numerical DMHs at various time are
illustrated in Fig.\ref{nhd}.  Based on this figure, we are
convinced that the numerical DMHs are sufficiently stable for
detecting significant DMH response to SSCs sinkage.

\subsection{Dynamical friction}

\begin{figure} 
      \vspace{-10.5 cm}
   \includegraphics[width=16cm]{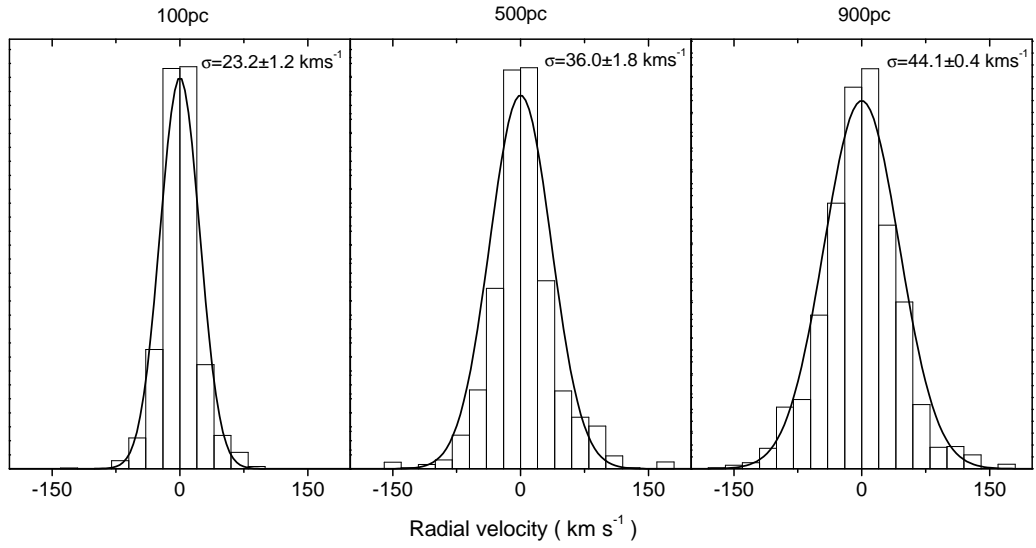}
      \vspace{-1.5 cm}
      \caption{Mean radial velocity distribution over the numerical unpolluted
      DMHs. Also shown are Gaussian distributions with  respective standard
deviations.} \label{nhvall}
   \end{figure}

In agreement with Kazantzidis et al. (2004), Fig.\ref{nhvall} shows
the one dimensional velocity distributions of the numerical DMHs. We
can see that the velocity distributions at distances not far from
the center are significantly steeper than Gaussian distribution.
According to Liu et al. (2005), this could make it inadequate to
estimate the dynamical friction with the formula we used in FHD03,
HDF03 and FHD04 (see also Binney \& Tremaine 1987; El-Zant et al.
2001). Instead, the \emph{Chandrasekhar dynamical friction formula},
valid for any isotropic velocity distribution, should be used
(Chandrasekhar 1943; Binney \& Tremaine 1987). This latter formula
reads
\begin{equation}
\label{Chdf}
\begin{array}{l}
\frac{d\vec{V}_M}{dt} = - 16 \pi^2 \ln\Lambda G^2 m (M+m) \frac{\int_0^{|\vec{V}_M|} f(v_m)v_m^2 dv_m}{|\vec{V}_M|^3} \vec{V}_M  \\
\hspace{0.8cm} \approx - 4 \pi \ln\Lambda G^2 M
\frac{\rho(|\vec{V}_M|)}{|\vec{V}_M|^3} \vec{V}_M
\end{array}
\end{equation}
where $f(v_m)$ is the phase-space number density of the background
composed of particles of mass $m$, moving with speed $v_m$; $M
(>>m)$ and $\vec{V}_M$ the mass and velocity of the cluster
experiencing the dynamical friction, respectively; $\ln\Lambda$ the
Coulomb logarithm; and $\rho(|\vec{V}_M|)$ denotes the mass density
of background matter moving slower than the cluster.

On the other hand, the dynamical friction of the stripped stars from
SSCs is roughly estimated using Maxwellian approximation, as we did
in our previous simulations.

\subsection{DMH responses}

There are two competing effects on the numerical DMH in response to
SSCs sinkage, namely the gravitational contraction and heating
expansion. The former one is automatically included in the
above-mentioned force model. We believe that the way to account for
the halo heating by El-Zant, Shlosman, \& Hoffman (2001) and El-Zant
et al. (2004), is a reasonable approach for us to take, because the
densely distributed DMH matter should also be efficient in
distributing, at least locally as is required, the energy obtained
from the coupling with the SSCs.

In practice, at the end of each integration step, ``the Cartesian
velocity components of the DMH particles are updated through an
additive term chosen from a normal distribution with zero mean and
variance'' (El-Zant, Shlosman, \& Hoffman 2001) $2E_b/3m$, where
$E_b$ is the energy gained by a DMH particle from all SSCs and $m$
the mass of the particle. We know that an SSC in its way spiraling
inwards will lose orbital energy mainly to DMH particles nearby its
trajectory. And as these particles with isotropic velocities
radially oscillate around their respective mean distances from the
galactic center, the energy lost by the SSC will be redistributed
among all DMH particles around $\bar{r}$,  where $\bar{r}$ is the
averaged distance of the SSC over the last integration step. This
implies that DMH particles around $\bar{r}$ should gain more energy
than the ones far from it. Based on the above arguments, we
distribute the energy lost by an SSC during the last integration
step to several bins, within each of which the particles (including
the involved DM not actually represented by particles. The
corresponding number of "particles" accounts for this part of DM can
be calculated from the fixed NFW density profile) will be assigned
with the same energy increment due to the considered SSC. These
SSC-dependent bins are obtained by dividing the radial distance
interval $[\max(0,\bar{r}-b),\bar{r}+b]$, where $b=2\ kpc$ the
effective impact parameter used in estimating the Coulomb logarithm.
The bin nearest to the SSC always covers the region explored by the
SSC in the last integration step, and the boundaries of the other
bins equispaced in $\log d$, where $d$ is the radial distance from
the SSC. The energy lost by an SSC is distributed among the bins
according to the following fact: the dynamical friction from the DMH
content within the distance $D$ from the SSC is proportional to $\ln
(1+ \Lambda_D^2)$, where $\Lambda_D \approx
\frac{D|\vec{V}_M|^2}{GM}$ (e.g. Binney \& Tremaine 1987).

It should be pointed out that the above simplifying semi-analytic
prescription for halo heating effectively neglects possible non
spherically symmetric evolution of the DMH, which can only be
studied with full N-body simulation.

\section{DMH evolution}
\begin{figure*} 
   \vspace{3cm}
   \hspace{0 cm}\includegraphics[width=18cm]{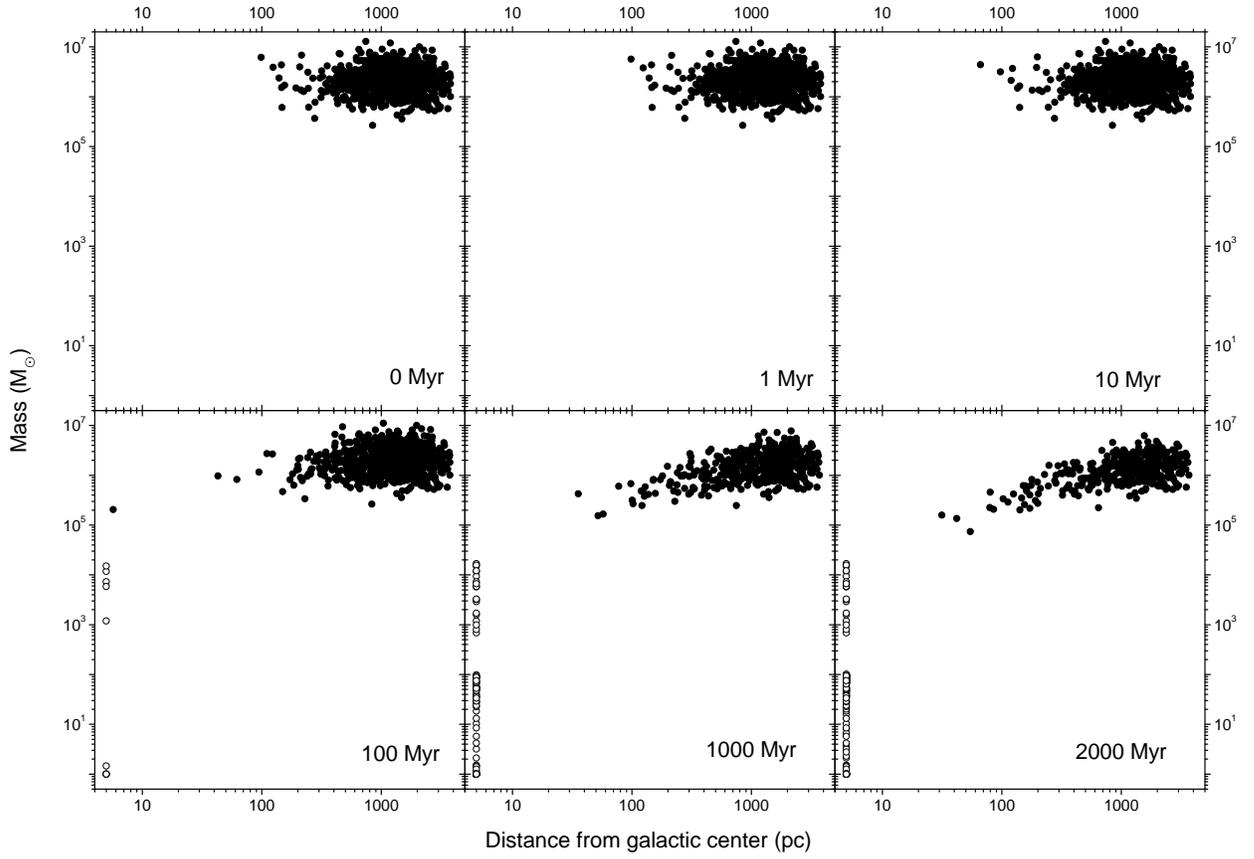}
      \vspace{-6cm}\caption{All SSCs in the 40 sample galaxies: mass vs distance from the center.
      Open circles indicate SSCs interior to 5 pc from the galactic center.} \label{2Sq}
   \end{figure*}

\begin{figure*} 
   \vspace{3cm}
   \hspace{0 cm} \includegraphics[width=16cm]{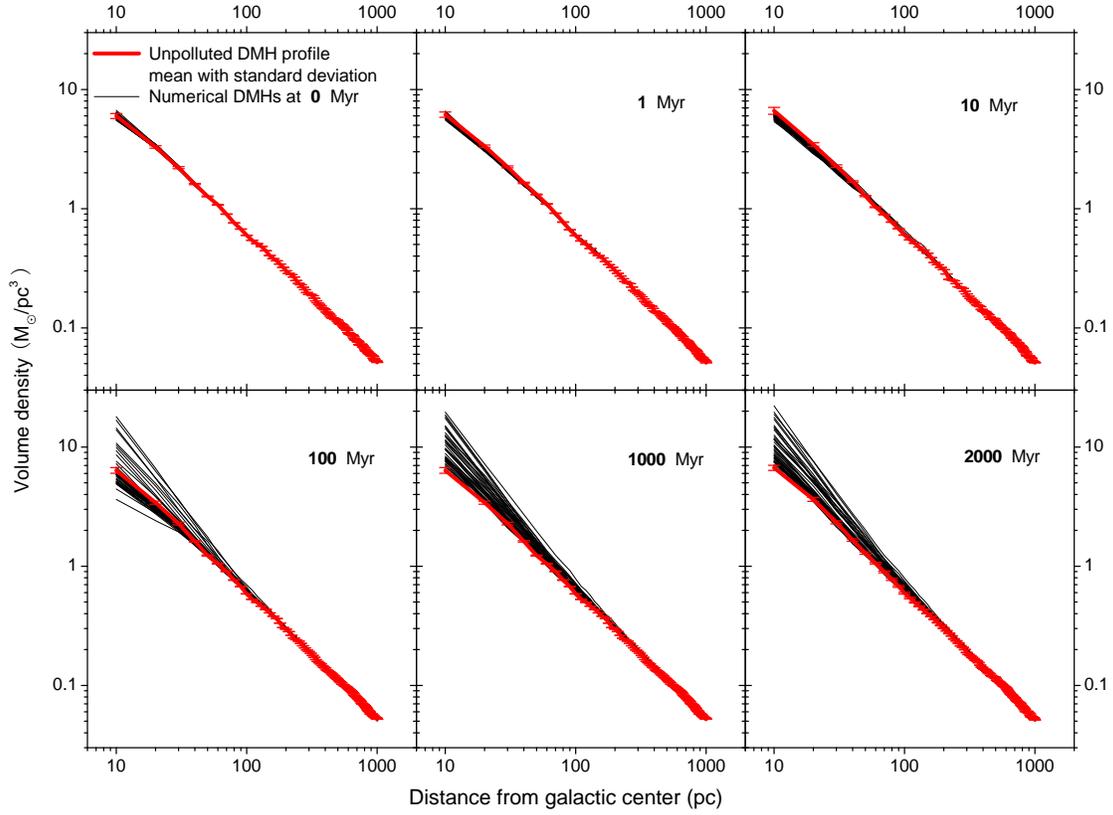}
      \vspace{-4 cm}\caption{Evolution of DMH density profiles induced by SSCs
sinkage.} \label{hd}
   \end{figure*}

For simplicity, the DMH used in this work will be referred to as the
responding DMH, while the DMH with NFW density profile and
Maxwellian velocity distribution used in HDF03  the fixed DMH.  In a
responding DMH, the evolution of DM density profile depends on the
redistribution of luminous matter induced by sinking of SSCs. Fig.
\ref{2Sq}  and Fig. \ref{hd} illustrate, respectively, the variation
of SSCs radial distribution with time and the resulting evolution of
DM density profiles.

The upper panels of Fig. \ref{2Sq} depict the early stage of SSCs'
sinkage until 10 Myrs, where we can find very few SSCs deposited to
the central region.  At that time, heating the halo induced by
dissipated orbital energy of spiraling inward of SSCs dominates. The
DM in the central region has been puffed outward then (also see,
Tasitsiomi 2003), leading to decreasing density in region of around
several tens of pc from the center. The upper panels of Fig.
\ref{hd} clearly present the early decreasing phase of this
 kind.  With the time going on, more and more SSCs have sunk into
nuclear region, shown in the lower panels of Fig. \ref{2Sq}. The
effect of gravitational contraction overwhelms that
 of heating expansion then, resulting in increasing DM distribution
 over that region. The steeper central cusps with respect to typical
NFW profiles are obviously indicated in the lower panels of Fig.
\ref{hd}.

\begin{figure*} 
   \vspace{3cm}
   \hspace{0 cm}\includegraphics[width=18cm]{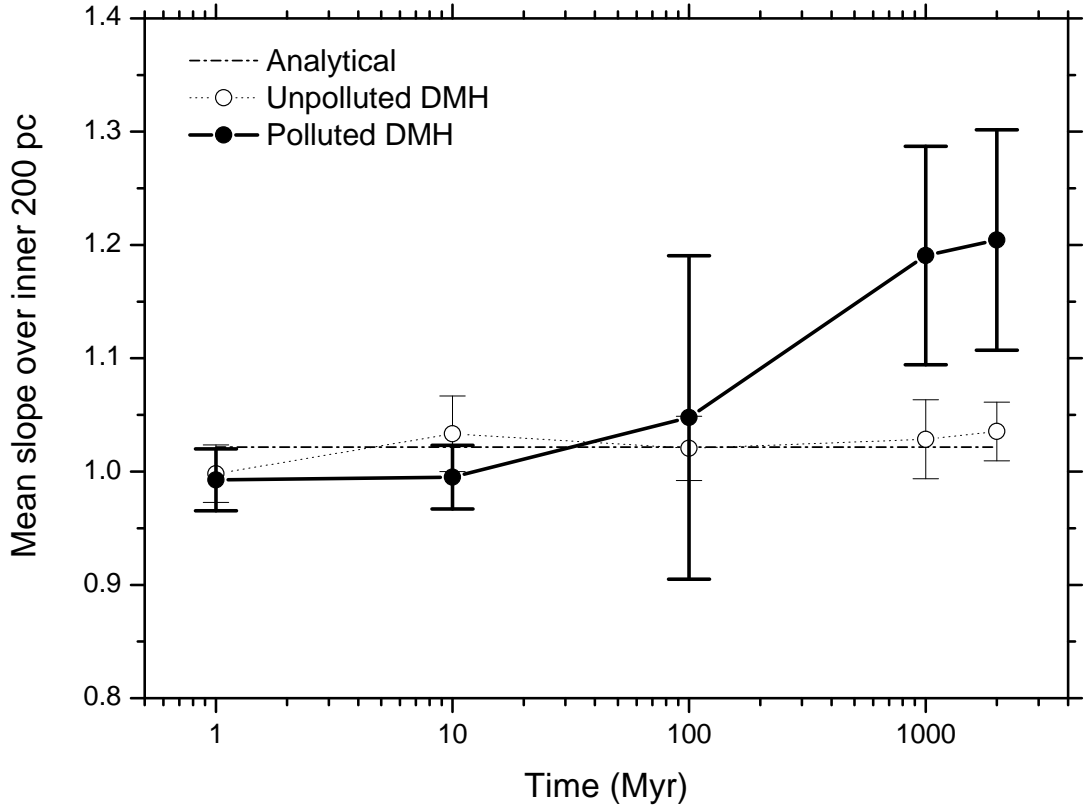}
      \vspace{-6cm}\caption{Evolution of mean inner slope of DMH density profile
      over inner $200\ pc$ globe. The notation of polluted
      DMH is used to indicate a DMH of which the mass
has been redistributed due to the influence of sinking SSCs.}
\label{sle}
\end{figure*}

We have noticed  that the DMH response is detectable only in region
less than $\sim$ 200pc
 from the center.  Adopting the logarithmic slope defined as
$\beta(r)=-{d\log(\rho)}/{d\log(r)}$  (Navvaro et al. 2004),  we
can show the time evolution of the mean responding DMH in Fig.
\ref{sle} by use of mean logarithmic slope over the inner 200pc
region. For comparison, the logarithmic slope of the analytical NFW
and the mean unpolluted DMH profiles are also indicated in this
figure. What we can see from this figure is an early decreasing
phase on the inner slope of the mean responding DM profile, followed
by gradual increasing trend with time. At 2 Gyrs, the inner slope of
the mean responding DM profile is about 1.2. A similar phenomenon
was recently obtained by Lin et al. (2006), who found that the
concentration parameters of DMHs increase  due to the influence of
baryonic matter on them.

\section{Bulge formation}

\subsection{Feedback of DMH response on bulge formation}
We have shown in the above section that the DM distribution
experiences two different phases induced by the influence of sinking
SSCs, the early decreasing and latter increasing density over the
inner central region. In return, the time evolution of DMH will
certainly affect the dynamical evolution of SSCs and the
bulge-formation histories.  To compare our present simulations (in a
responding DMH) with those  in the fixed DMH
 (HDF03) would clarify the feedback of DMH response on bulge formation, shedding
light on galaxy formation.

\begin{figure*} 
   \vspace{3cm}
   \hspace{0 cm}\includegraphics[width=18cm]{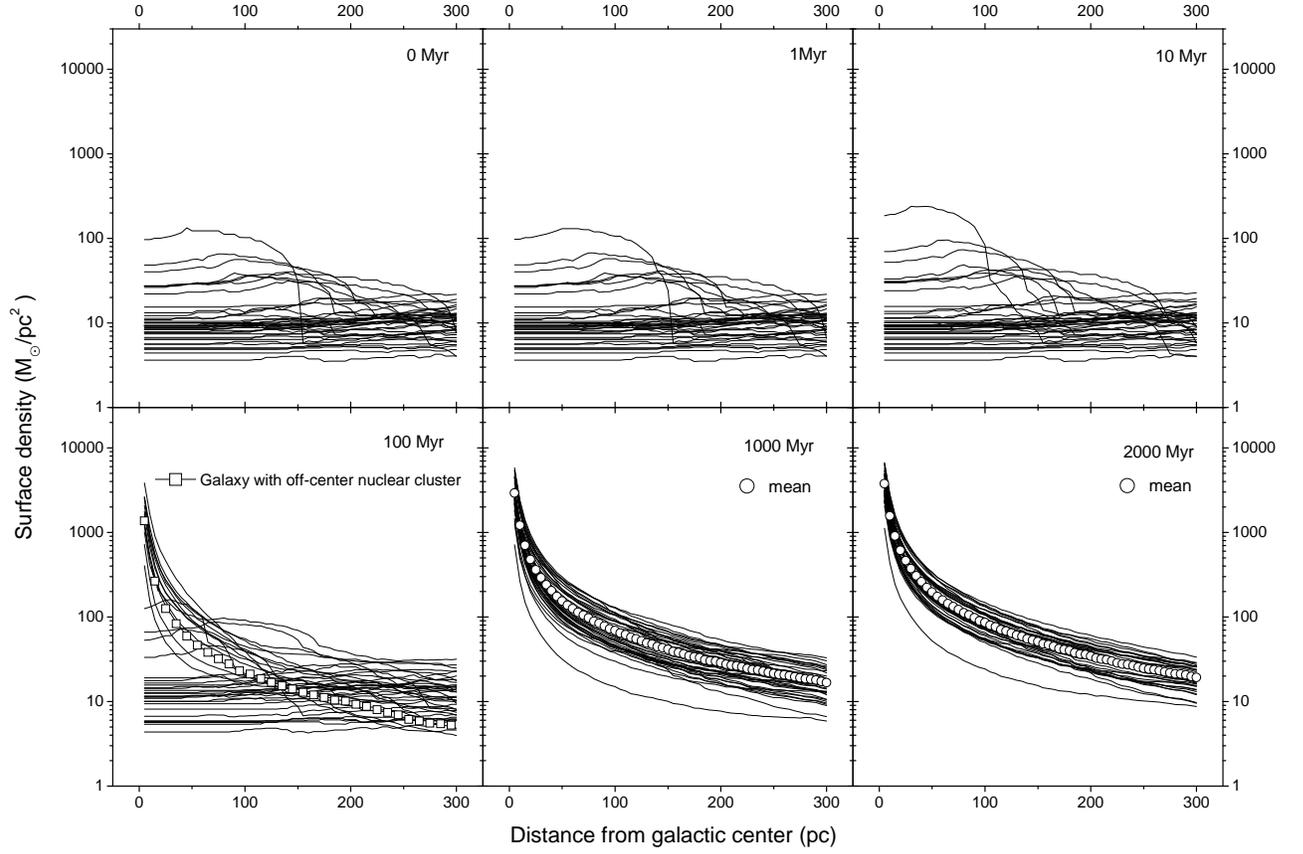}
      \vspace{-6cm}\caption{Evolution of surface density profiles of luminous matter.}
\label{bsd}
   \end{figure*}

\begin{figure*} 
   \vspace{3cm}
   \hspace{0 cm}\includegraphics[width=18cm]{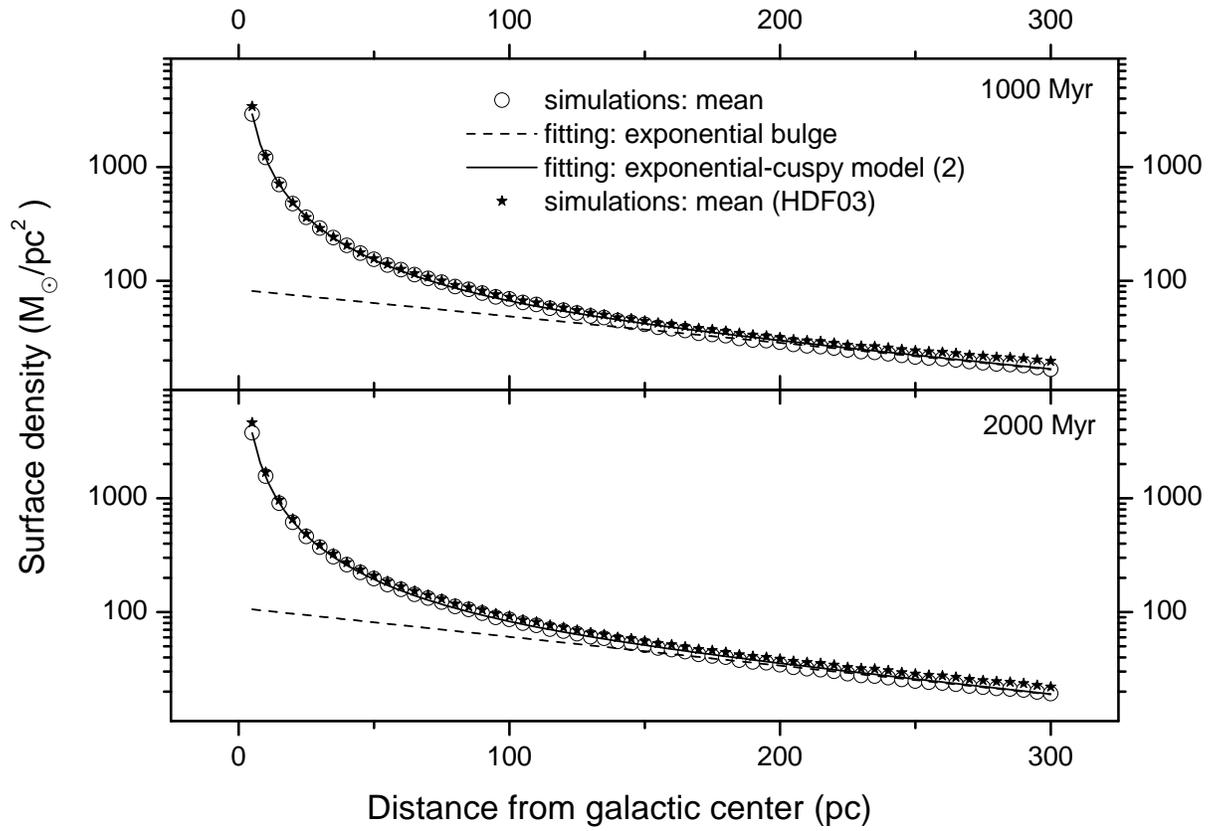}
      \vspace{-6 cm} \caption{Mean surface density profiles of the formed bulges.}
\label{bsdfit}
   \end{figure*}

\begin{figure*} 
   \vspace{3cm}
   \hspace{0 cm} \vspace{-6cm} \includegraphics[width=18cm]{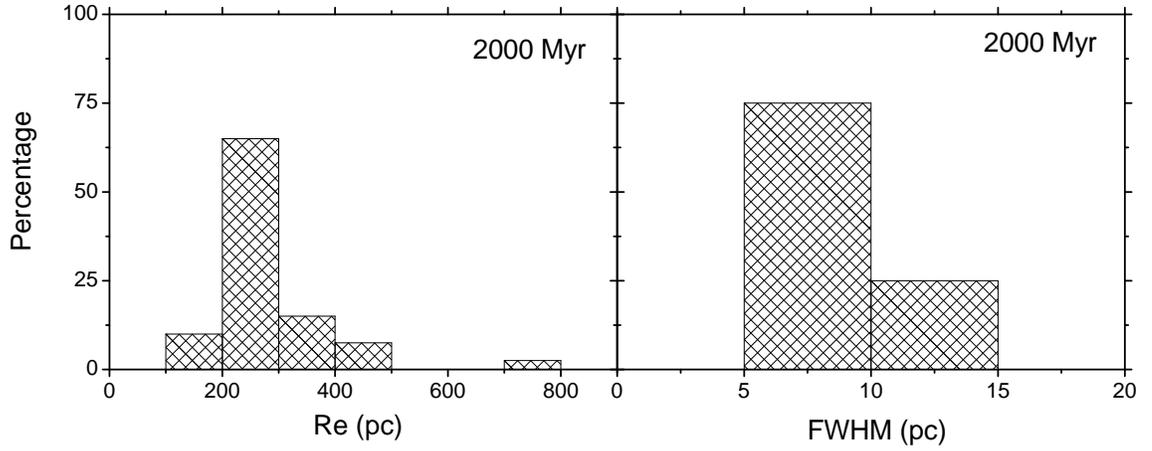}
      \vspace{0cm}\caption{Histograms of scale lengths of the formed bulges, $R_e$
      for the outer exponential component,
      and FWHM for the inner cuspy component. The linear resolution of $5\ pc$ is
      used to derive FWHM.} \label{ReFWHM}
   \end{figure*}

Following what we did in the fixed DMH (HDF03), we display the bulge
formation processes and the mean surface density profiles in Fig.
\ref{bsd}  and Fig. \ref{bsdfit}, respectively. In general, they are
about the same as what we obtained in HDF03, i.e. the formed bulges
are characterized by the general presence of a central cusp on top
of an exponential bulge. For this reason, we take the same fitting
models as we used in HDF03 (Carollo \& stiavalli 1998)

\begin{equation} \label{model}
\sigma(R)=\sigma_0 \exp(-1.678
\frac{R}{R_e})+\sigma_1(1+\frac{R_c}{R})^\gamma
\exp(-\frac{R}{R_s}),
\end{equation}

\noindent where the first term represents an exponential bulge and
the second one an additional nuclear cusp. As shown by open circles
in Fig. \ref{bsdfit}, the mean profiles obtained from the present
simulations are well fitted to this model. The solid stars showing
in Fig. \ref{bsdfit} denote the derived mean profiles in HDF03. They
are just slightly higher than the formed bulges in the responding
DMH. It follows that the histogram of the scale length, $R_e$, for
each simulated bulges and FWHM for nuclear cusps derived in
responding DMH, displayed in Fig. \ref{ReFWHM}, would be about the
same as those obtained in fixed DMH (HDF03), though the later
results are acquired at age of 3 Gyrs.

The disparity, though small, of mean surface density profiles in
different DMHs  should have something to do with the feedback of the
DMH response on bulge formation. In fact, most SSCs take more than
several hundred Myrs to spiral down to several hundred pc from the
center, shown in Fig. \ref{2Sq}. On the other hand, apart from the
very early time the DMH response leads to increasing density of DM
over the central region. It would cause stronger tidal stripping on
SSCs outside the bulge region, resulting in lower surface density
distribution than that derived in the case of fixed DMH, just looks
like what we illustrated in Fig. \ref{bsdfit}.

The continuously stronger tidal stripping induced by DMH response in
most of evolution time would make it harder for an SSC to survive as
an off-center star cluster than in the fixed DMH. Nonetheless, one
such cluster presents at about 6 pc from the center in the snapshot
at $100\ Myr$ in Fig. \ref{2Sq} . The surface density profile of the
bulge hosting this cluster is indicated by squares in the lower left
panel of Fig. \ref{bsd}. Clearly, the outer part of this bulge is
very weak. This simulated sample provides a possible explanation for
the existence of off-center nuclear star clusters hosted in the so
called bulgeless spirals (Matthews \& Gallagher 2002). Indeed,
observations on searching for bulges in very late-type galaxies
started only very recently, and it's possible that some very weak
bulges remain undetected (see, e.g. B\"{o}ker et al. 2003; Carollo
1999).

\begin{figure*} 
   \vspace{3cm}
   \hspace{0 cm}\includegraphics[width=18cm]{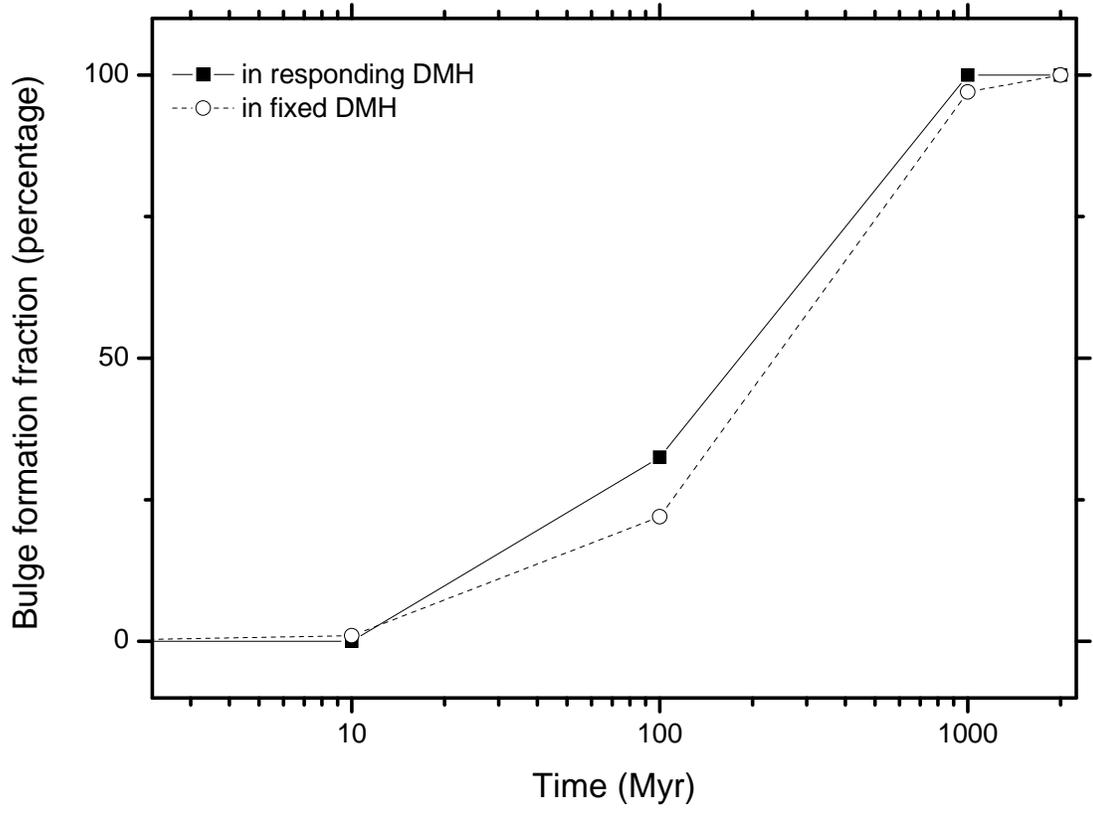}
      \vspace{-6cm}\caption{Bulge formation fraction.}
\label{formationrate}
   \end{figure*}

Fig. \ref{formationrate} presents the bulge formation fractions in
the responding and the fixed DMHs. Generally speaking, bulges form
earlier in the former DMH. This is mainly resulted form its steeper
velocity distribution than the Gaussian one. In fact, for two DMHs
having the same volume density, the one with steeper velocity
distribution has stronger effect of dynamical friction (Binney \&
Tremaine 1987; Liu et al. 2005). Besides, we have discussed above
that the DMH density response should also be in favor of stronger
effect of dynamical friction on most SSCs.  However, the opposite
situation is possible for SSCs initially very close to the central
region where DMH density undergoes a transitory, early decreasing
phase, seen in Fig. \ref{hd}. This explains why the bulge formation
fraction at very early stage is lower in the responding DMH than in
the fixed DMH .

In summary,  the DMH response to SSCs sinkage does have detectable
feedbacks on  bulge formation. However, the feedbacks discussed
above are actually not very significant. The formed bulges in both
fixed and responding DMHs are basically the same at least in the
view of the present day observations. It suggests that the fixed DMH
could still be an acceptable simplifying model for dark matter of
galaxies, as far as the bulge formation is concerned.

\subsection{Mass correlation between the paired bulges and nuclear cusps}
The occurrence of compact massive star clusters at nuclei of spiral
galaxies is now believed to be a common phenomenon (e.g. B\"{o}ker
et al. 2002, Matthews \& Gallagher 2002 ; B\"{o}ker et al. 2003;
Rossa et al. 2006). The great  progress made by Rossa et al. (2006)
is to derive important properties of these nuclear clusters (NCs),
such as ages, masses, and their composition of stellar populations.
According to these investigations, we have a basic understanding to
the physical properties of NCs hosted by most spirals. For example,
the NC effective radii are typically in the range of 2 - 10 pc,
their average masses ($\overline{\log M}$) and mass-weighted ages
($\overline{<log \tau>}_{M}$) are, respectively, 6.25 vs 7.63 and
9.07 vs 9.89  for late- vs early-type spirals.

The most instructive result by Rossa et al. (2006) is the tight
correlation of the NC mass
 to the luminosity of its bulge, similar to the striking luminosity correlation of
paired bulge and the nuclear cusp obtained by Balcells et al.
(2003). Both of these studies  strongly imply the causally connected
formation processes between the two components. In our previous work
in the fixed DMH (HDF03), we have shown clearly that the mass
relation between the simulated bulges and their paired nuclear
(stellar) cusps coincides well with  what Balcells et al. (2003)
obtained.

\begin{figure*} 
   \vspace{3cm}
   \hspace{0 cm}\includegraphics[width=18cm]{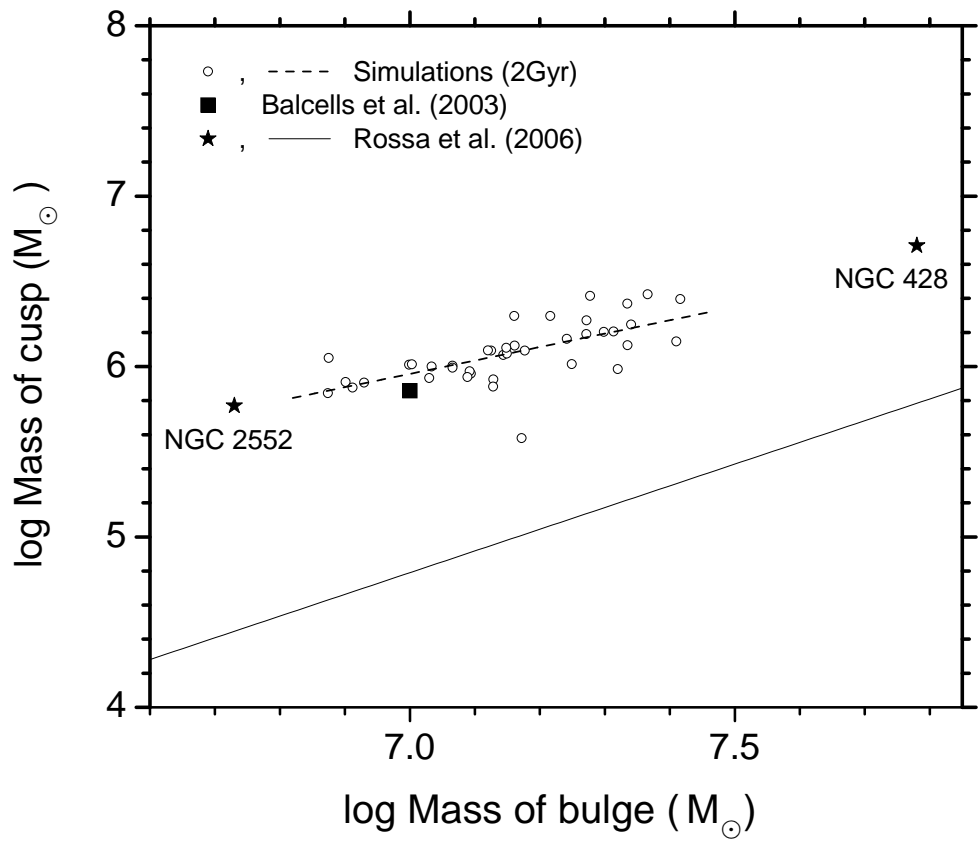}
      \vspace{-6cm}\caption{Correlation between the masses of bulge and it's paired cusp.
      This figure shows that our simulations for spirals experiencing one starburst event
      is in good agreement with observations for very late-type spirals (NGC 428 and
      NGC2552 (Rossa et al., 2006)). The disparity between our simulated dashed line and
      the observed thin line (Rossa et al., 2006) results from the fact that most sample
      galaxies are not very late-type ones.}
\label{bcmHD}
  \end{figure*}

In the case of responding DMH, the mass relation between the nuclear
(stellar) cusps and their paired bulges would also hold as expected
from the connected formation processes of these two components,  as
shown in Fig. \ref{bcmHD}.  The dashed line in the figure
illustrates the linear fitting for these simulated data. The data
point, in solid square, derived for Balcells et al's situation (see,
HDF03) is also added in this figure for comparison. The thin solid
line is extracted from Rossa et al. (2006), based on certain
mass-to-light ratio.

The major problem for comparing our simulations to what Rossa et al
(2006) derived lies in their important finding, i.e. NCs are
composed of mixed populations of different ages. It means that NCs
form via at least more than one starburst event. On the contrary,
what we proposed to form bulges and nuclear (stellar) cusps through
dynamical evolution of SSCs in DM dominated galaxies is only for
very late-type spirals experiencing one starburst event, triggered
by galactic harassment. This inherent difference causes the
disparity of our simulations from the observed thin line shown in
Fig. \ref{bcmHD}.

According to the hierarchical processes of galaxy formation, the
very late-type spirals would be the ones undergoing least
merger/interaction events. That is to say, two Sm type spirals among
the 40 sample galaxies that Rossa et al. (2006) observed, NGC 428
and NGC2552, are sources  most suitable for comparing with our
simulations. The solid stars shown in Fig. \ref{bcmHD} are the data
points of these two galaxies. Though two sample galaxies  are really
not enough for us to reach definite conclusion, the close slopes of
the two (observed) data points and the dashed (simulation) lines is
very encouraging indeed.

\begin{figure*} 
    \hspace{0 cm} \includegraphics[width=18cm]{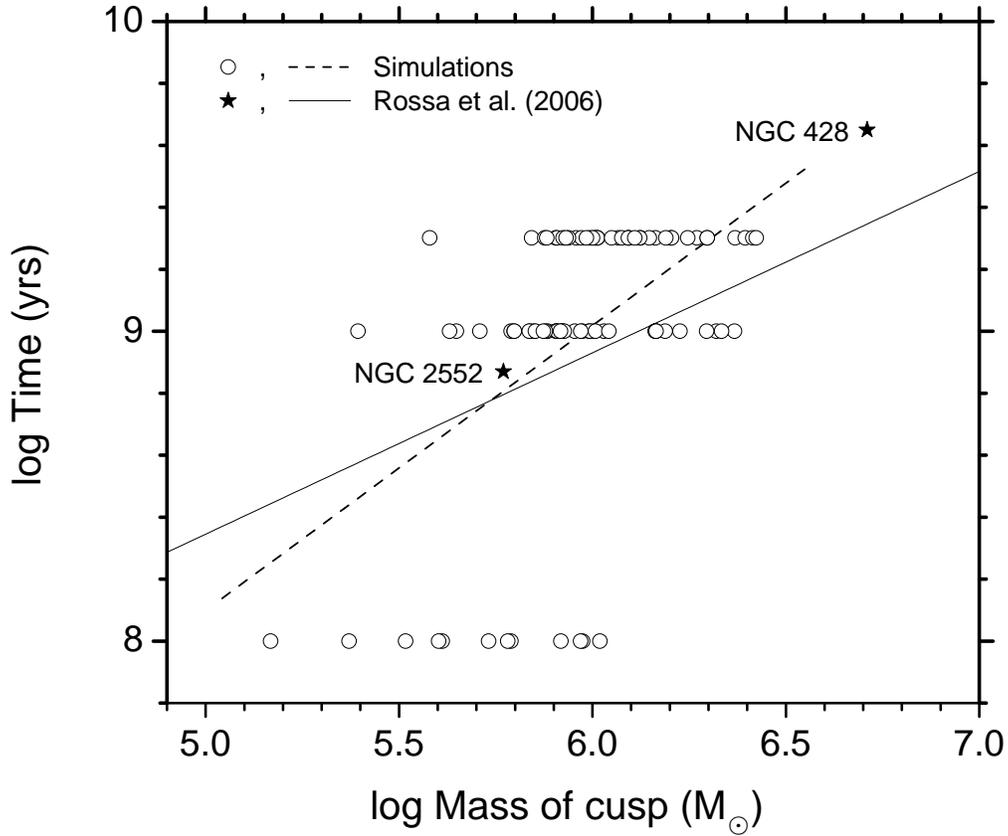}
      \vspace{-6cm}\caption{Correlation between the mass and age of the nuclear cusp.
      Exactly the same situation as the correlation between the masses of bulge and
      it's paired cusp (see Fig.10), our simulations for spirals experiencing one
      starburst event is in good agreement with observations only for very late-type
      galaxies (NGC 428 and NGC2552 (Rossa et al., 2006)).}
\label{mtcmHD}
   \end{figure*}

The same situation as we see in Fig. \ref{bcmHD}  also occurs in the
time dependency of nuclear (stellar) cusp mass, shown in Fig.
\ref{mtcmHD}. The thin solid line extracted from Rossa et al.(2006)
illustrates the NC mass in relation to the mass-weighted ages. The
mass-weighted ages are referred to  the older population of NCs,
which contains most of the NCs mass. Obviously, it is this case that
does match what our simulations derive. The most impressive thing in
this figure is the very close slopes of our simulations to the two
galaxies with Hubble type of Sm, NGC 428 and NGC2552, which is just
the same as we see in Fig. \ref{bcmHD}.

The same closeness appearing in two different  statistic
correlations gives us a very strong hint that our hypothesis for
bulge/(nuclear cusps) formation does  have something to do with NC
formation. In fact, our previous simulations with higher resolution
of 2pc (HDF03) showed that the effective radii of nuclear (stellar)
cusps range from a few - 9 pc, and their median mass $1.5 \times
10^{6} M_\odot$ mass, both of them are in accordance with the
observed data mentioned above. In combination of all of these
arguments,  we are convinced that the hypothesis that we proposed
(FHD03) could be a very promising mechanism for NC formation in most
spiral galaxies, apart from being an approach to form bulges.

\section{Conclusions}

The increasingly observed data indicate that galactic harassments,
even minor mergers, such as the case of NGC 3310 (de Grijs et al.
2003), can trigger the formation of a set of SSCs over galaxies. The
follow-up, dynamical evolution of SSCs in a configuration of DM
dominated systems would be a general phenomenon, which motivated our
investigation on this processes to be a model for bulge formation
(FHD03).

As a basic part of this model, the evolution of DMH induced by the
influence of sinking SSCs was a major step that we moved on. It is
also of interest to understand the resulting variation of the inner
slopes of the DM density profiles in that case. The simulations
performed in this work have indicated that the DMH response to
sinking of SSCs does cause the inner slope of an initial NFW density
profile to steepen, approaching to 1.2 at 2 Gyrs in mean logarithmic
slope over the responding region.

However, compared with the region traversed by sinking SSCs, the
above mentioned responding region is small. As a result, the
feedbacks of the DMH response on the bulge formation turned out to
be very small, in the sense that the formed bulges in both the fixed
(HDF03) and the responding DMH (this work) are basically the same,
both are consistent with $HST$ observations.

One very instructive result obtained from our consecutive
investigations (FHD03; HDF03; and this work) is that nuclear
(stellar) cusps are formed, no matter what kind of DM density
profile was adopted, the NFW or the Burkert profile, no matter
whether the interaction between the dark and luminous matter has
been considered or not. The derived median mass and effective radii
of the cusps are accordance with the observed corresponding NC
quantities in most spiral galaxies.

The more important point on this matter would be what we
demonstrated in this work on the mass correlation of nuclear
(stellar) cusps to the bulges, which is consistent with the similar
observed relation for relevant sample spirals. The same situation
appears in diverse statistics on the time dependency of NC/(nuclear
cusp) mass.

No matter how complicated the formation processes are for nuclear
clusters of spirals, the work that we have done indicates that to
form nuclear cusps through sinking of SSCs would be a very promising
mechanism for NC formation in very late-type spirals. Also, they
could be the base, or seed, to grow up for late-type spirals that
Rossa et al. (2006) observed.

Obviously, more work needs to be done in both observations and
simulations so that we could reach positive conclusions. Especially,
to observe more galaxies with Hubble type of Sm and to simulate
bulge formation or growth originated from more than one galactic
harassment event would be desperately needed.

\acknowledgements
The anonymous referee is thanked for his critical
comments, which clarify and strengthen the reasoning of the paper.
This work is supported by NSFC 10373008. Fu and Deng are also
supported by NSFC key program 10233020 and 10333060, respectively.

\clearpage
\begin{deluxetable}{lr}
\tablecolumns{2}
\tablewidth{0pc}
\tablecaption{Models}
\tablehead{
 \colhead{Item} & \colhead{Description} }
\startdata

DM DMH: initial density profile & NFW with $M_{200}=10^{11} M_{\odot}$ \\
SSC: initial mass function & Log-normal with \\
& $\exp(mean)=2 \times 10^6 M_{\odot}, variance=0.08$ \\
SSC: projected radial number distribution & $N(R) \sim R{\hspace*{0.6mm}} \exp(-R/0.6\ kpc)$ \\
SSC: total number in a single galaxy & 17 \\
SSC: initial velocity & Local circular speed
\enddata
\end{deluxetable}

\clearpage

\clearpage

\clearpage

\clearpage

\clearpage

\clearpage

\clearpage

\clearpage

\clearpage


\clearpage

\clearpage

\end{document}